
\font \tbfontt                = cmbx10 scaled\magstep1
\font \tafontt                = cmbx10 scaled\magstep2
\font \tbfontss               = cmbx5  scaled\magstep1
\font \tafontss               = cmbx5  scaled\magstep2
\font \sixbf                  = cmbx6
\font \tbfonts                = cmbx7  scaled\magstep1
\font \tafonts                = cmbx7  scaled\magstep2
\font \ninebf                 = cmbx9
\font \tasys                  = cmex10 scaled\magstep1

\font \sixi                   = cmmi6
\font \ninei                  = cmmi9
\font \tams                   = cmmib10
\font \tbmss                  = cmmib10 scaled 600
\font \tamss                  = cmmib10 scaled 700
\font \tbms                   = cmmib10 scaled 833
\font \tbmt                   = cmmib10 scaled\magstep1
\font \tamt                   = cmmib10 scaled\magstep2
\font \smallescriptscriptfont = cmr5
\font \smalletextfont         = cmr5 at 10pt
\font \smallescriptfont       = cmr5 at 7pt
\font \sixrm                  = cmr6
\font \ninerm                 = cmr9
\font \ninesl                 = cmsl9
\font \tensans                = cmss10
\font \fivesans               = cmss10 at 5pt
\font \sixsans                = cmss10 at 6pt
\font \sevensans              = cmss10 at 7pt
\font \ninesans               = cmss10 at 9pt
\font \tbst                   = cmsy10 scaled\magstep1
\font \tast                   = cmsy10 scaled\magstep2
\font \tbsss                  = cmsy5  scaled\magstep1
\font \tasss                  = cmsy5  scaled\magstep2
\font \sixsy                  = cmsy6
\font \tbss                   = cmsy7  scaled\magstep1
\font \tass                   = cmsy7  scaled\magstep2
\font \ninesy                 = cmsy9
\font \markfont               = cmti10 at 11pt
\font \nineit                 = cmti9
\font \ninett                 = cmtt9
\magnification=\magstep0
\hsize=13truecm
\vsize=19.8truecm
\hfuzz=2pt
\tolerance=500
\abovedisplayskip=3 mm plus6pt minus 4pt
\belowdisplayskip=3 mm plus6pt minus 4pt
\abovedisplayshortskip=0mm plus6pt minus 2pt
\belowdisplayshortskip=2 mm plus4pt minus 4pt
\predisplaypenalty=0
\clubpenalty=10000
\widowpenalty=10000
\frenchspacing
\newdimen\oldparindent\oldparindent=1.5em
\parindent=1.5em



\def\bbbc{{\mathchoice {\setbox0=\hbox{$\displaystyle\rm C$}\hbox{\hbox
to0pt{\kern0.4\wd0\vrule height0.9\ht0\hss}\box0}}
{\setbox0=\hbox{$\textstyle\rm C$}\hbox{\hbox
to0pt{\kern0.4\wd0\vrule height0.9\ht0\hss}\box0}}
{\setbox0=\hbox{$\scriptstyle\rm C$}\hbox{\hbox
to0pt{\kern0.4\wd0\vrule height0.9\ht0\hss}\box0}}
{\setbox0=\hbox{$\scriptscriptstyle\rm C$}\hbox{\hbox
to0pt{\kern0.4\wd0\vrule height0.9\ht0\hss}\box0}}}}
\def\bbbe{{\mathchoice {\setbox0=\hbox{\smalletextfont e}\hbox{\raise
0.1\ht0\hbox to0pt{\kern0.4\wd0\vrule width0.3pt
height0.7\ht0\hss}\box0}}
{\setbox0=\hbox{\smalletextfont e}\hbox{\raise 0.1\ht0\hbox
to0pt{\kern0.4\wd0\vrule width0.3pt height0.7\ht0\hss}\box0}}
{\setbox0=\hbox{\smallescriptfont e}\hbox{\raise 0.1\ht0\hbox
to0pt{\kern0.5\wd0\vrule width0.2pt height0.7\ht0\hss}\box0}}
{\setbox0=\hbox{\smallescriptscriptfont e}\hbox{\raise
0.1\ht0\hbox to0pt{\kern0.4\wd0\vrule width0.2pt
height0.7\ht0\hss}\box0}}}}
\def\bbbq{{\mathchoice {\setbox0=\hbox{$\displaystyle\rm Q$}\hbox{\raise
0.15\ht0\hbox to0pt{\kern0.4\wd0\vrule height0.8\ht0\hss}\box0}}
{\setbox0=\hbox{$\textstyle\rm Q$}\hbox{\raise
0.15\ht0\hbox to0pt{\kern0.4\wd0\vrule height0.8\ht0\hss}\box0}}
{\setbox0=\hbox{$\scriptstyle\rm Q$}\hbox{\raise
0.15\ht0\hbox to0pt{\kern0.4\wd0\vrule height0.7\ht0\hss}\box0}}
{\setbox0=\hbox{$\scriptscriptstyle\rm Q$}\hbox{\raise
0.15\ht0\hbox to0pt{\kern0.4\wd0\vrule height0.7\ht0\hss}\box0}}}}
\def\bbbt{{\mathchoice {\setbox0=\hbox{$\displaystyle\rm
T$}\hbox{\hbox to0pt{\kern0.3\wd0\vrule height0.9\ht0\hss}\box0}}
{\setbox0=\hbox{$\textstyle\rm T$}\hbox{\hbox
to0pt{\kern0.3\wd0\vrule height0.9\ht0\hss}\box0}}
{\setbox0=\hbox{$\scriptstyle\rm T$}\hbox{\hbox
to0pt{\kern0.3\wd0\vrule height0.9\ht0\hss}\box0}}
{\setbox0=\hbox{$\scriptscriptstyle\rm T$}\hbox{\hbox
to0pt{\kern0.3\wd0\vrule height0.9\ht0\hss}\box0}}}}
\def\bbbs{{\mathchoice
{\setbox0=\hbox{$\displaystyle     \rm S$}\hbox{\raise0.5\ht0\hbox
to0pt{\kern0.35\wd0\vrule height0.45\ht0\hss}\hbox
to0pt{\kern0.55\wd0\vrule height0.5\ht0\hss}\box0}}
{\setbox0=\hbox{$\textstyle        \rm S$}\hbox{\raise0.5\ht0\hbox
to0pt{\kern0.35\wd0\vrule height0.45\ht0\hss}\hbox
to0pt{\kern0.55\wd0\vrule height0.5\ht0\hss}\box0}}
{\setbox0=\hbox{$\scriptstyle      \rm S$}\hbox{\raise0.5\ht0\hbox
to0pt{\kern0.35\wd0\vrule height0.45\ht0\hss}\raise0.05\ht0\hbox
to0pt{\kern0.5\wd0\vrule height0.45\ht0\hss}\box0}}
{\setbox0=\hbox{$\scriptscriptstyle\rm S$}\hbox{\raise0.5\ht0\hbox
to0pt{\kern0.4\wd0\vrule height0.45\ht0\hss}\raise0.05\ht0\hbox
to0pt{\kern0.55\wd0\vrule height0.45\ht0\hss}\box0}}}}
\def\bbbz{{\mathchoice {\hbox{$\sans\textstyle Z\kern-0.4em Z$}}
{\hbox{$\sans\textstyle Z\kern-0.4em Z$}}
{\hbox{$\sans\scriptstyle Z\kern-0.3em Z$}}
{\hbox{$\sans\scriptscriptstyle Z\kern-0.2em Z$}}}}
\skewchar\ninei='177 \skewchar\sixi='177
\skewchar\ninesy='60 \skewchar\sixsy='60
\hyphenchar\ninett=-1
\def\newline{\hfil\break}%
\catcode`@=11
\def\folio{\ifnum\pageno<\z@
\uppercase\expandafter{\romannumeral-\pageno}%
\else\number\pageno \fi}
\catcode`@=12 
  \mathchardef\Gamma="0100
  \mathchardef\Delta="0101
  \mathchardef\Theta="0102
  \mathchardef\Lambda="0103
  \mathchardef\Xi="0104
  \mathchardef\Pi="0105
  \mathchardef\Sigma="0106
  \mathchardef\Upsilon="0107
  \mathchardef\Phi="0108
  \mathchardef\Psi="0109
  \mathchardef\Omega="010A
\def\squareforqed{\hbox{\rlap{$\sqcap$}$\sqcup$}}
\def\qed{\ifmmode\squareforqed\else{\unskip\nobreak\hfil
\penalty50\hskip1em\null\nobreak\hfil\squareforqed
\parfillskip=0pt\finalhyphendemerits=0\endgraf}\fi}
\newfam\sansfam
\textfont\sansfam=\tensans\scriptfont\sansfam=\sevensans
\scriptscriptfont\sansfam=\fivesans
\def\sans{\fam\sansfam\tensans}
\def\stackfigbox{\if
Y\FIG\global\setbox\figbox=\vbox{\unvbox\figbox\box1}%
\else\global\setbox\figbox=\vbox{\box1}\global\let\FIG=Y\fi}
\def\placefigure{\dimen0=\ht1\advance\dimen0by\dp1
\advance\dimen0by5\baselineskip
\advance\dimen0by0.4true cm
\ifdim\dimen0>\vsize\pageinsert\box1\vfill\endinsert
\else
\if Y\FIG\stackfigbox\else
\dimen0=\pagetotal\ifdim\dimen0<\pagegoal
\advance\dimen0by\ht1\advance\dimen0by\dp1\advance\dimen0by1.7true cm
\ifdim\dimen0>\pagegoal\stackfigbox
\else\box1\vskip7true mm\fi
\else\box1\vskip7true mm\fi\fi\fi\let\firstleg=Y}
%
\def\begfig#1cm#2\endfig{\par
\setbox1=\vbox{\dimen0=#1true cm\advance\dimen0
by1true cm\kern\dimen0\vskip-.8333\baselineskip#2}\placefigure}
\def\begdoublefig#1cm #2 #3 \enddoublefig{\begfig#1cm%
\line{\vtop{\hsize=0.46\hsize#2}\hfill
\vtop{\hsize=0.46\hsize#3}}\endfig}
\let\firstleg=Y
\def\figure#1#2{\if Y\firstleg\vskip1true cm\else\vskip1.7true mm\fi
\let\firstleg=N\setbox0=\vbox{\noindent\petit{\bf
Fig.\ts#1\unskip.\ }\ignorespaces #2\smallskip
\count255=0\global\advance\count255by\prevgraf}%
\ifnum\count255>1\box0\else
\centerline{\petit{\bf Fig.\ts#1\unskip.\
}\ignorespaces#2}\smallskip\fi}

\def\begtab#1cm#2\endtab{\par
   \ifvoid\topins\midinsert\medskip\vbox{#2\kern#1true cm}\endinsert
   \else\topinsert\vbox{#2\kern#1true cm}\endinsert\fi}
\def\begpet{\vskip6pt\bgroup\petit}
\def\endpet{\vskip6pt\egroup}
\newdimen\refindent
\newlinechar=`\^
\def\begref#1#2{\titlea{}{#1}%
\bgroup\petit
\setbox0=\hbox{#2\enspace}\refindent=\wd0\relax
\if>#2>\else
\ifdim\refindent>0.5em\else
\message{^Something may be wrong with your references;}%
\message{probably you missed the second argument of \string\begref.}%
\fi\fi}
\def\ref{\goodbreak
\hangindent\oldparindent\hangafter=1
\noindent\ignorespaces}
\def\refno#1{\goodbreak
\setbox0=\hbox{#1\enspace}\ifdim\refindent<\wd0\relax
\message{^Your reference `#1' is wider than you pretended in using
\string\begref.}\fi
\hangindent\refindent\hangafter=1
\noindent\kern\refindent\llap{#1\enspace}\ignorespaces}
\def\refmark#1{\goodbreak
\setbox0=\hbox{#1\enspace}\ifdim\refindent<\wd0\relax
\message{^Your reference `#1' is wider than you pretended in using
\string\begref.}\fi
\hangindent\refindent\hangafter=1
\noindent\hbox to\refindent{#1\hss}\ignorespaces}
\def\endref{\goodbreak\endpet}
\def\petit{\def\rm{\fam0\ninerm}%
\textfont0=\ninerm \scriptfont0=\sixrm \scriptscriptfont0=\fiverm
 \textfont1=\ninei \scriptfont1=\sixi \scriptscriptfont1=\fivei
 \textfont2=\ninesy \scriptfont2=\sixsy \scriptscriptfont2=\fivesy
 \def\it{\fam\itfam\nineit}%
 \textfont\itfam=\nineit
 \def\sl{\fam\slfam\ninesl}%
 \textfont\slfam=\ninesl
 \def\bf{\fam\bffam\ninebf}%
 \textfont\bffam=\ninebf \scriptfont\bffam=\sixbf
 \scriptscriptfont\bffam=\fivebf
 \def\sans{\fam\sansfam\ninesans}%
 \textfont\sansfam=\ninesans \scriptfont\sansfam=\sixsans
 \scriptscriptfont\sansfam=\fivesans
 \def\tt{\fam\ttfam\ninett}%
 \textfont\ttfam=\ninett
 \normalbaselineskip=11pt
 \setbox\strutbox=\hbox{\vrule height7pt depth2pt width0pt}%
 \normalbaselines\rm
\def\vec##1{{\textfont1=\tbms\scriptfont1=\tbmss
\textfont0=\ninebf\scriptfont0=\sixbf
\mathchoice{\hbox{$\displaystyle##1$}}{\hbox{$\textstyle##1$}}
{\hbox{$\scriptstyle##1$}}{\hbox{$\scriptscriptstyle##1$}}}}}
\nopagenumbers
%
\let\header=Y
\let\FIG=N
\newbox\figbox
\output={\if N\header\headline={\hfil}\fi\plainoutput
\global\let\header=Y\if Y\FIG\topinsert\unvbox\figbox\endinsert
\global\let\FIG=N\fi}
\let\lasttitle=N
\def\centerpar#1{{\parfillskip=0pt
\rightskip=0pt plus 1fil
\leftskip=0pt plus 1fil
\advance\leftskip by\oldparindent
\advance\rightskip by\oldparindent
\def\newline{\break}%
\noindent\ignorespaces#1\par}}
\catcode`\@=\active
\def\author#1{\bgroup
\baselineskip=13.2pt
\lineskip=0pt
\pretolerance=10000
\markfont
\centerpar{#1}\bigskip\egroup
{\def@##1{}%
\setbox0=\hbox{\petit\kern2.5true cc\ignorespaces#1\unskip}%
\ifdim\wd0>\hsize
\message{The names of the authors exceed the headline, please use a }%
\message{short form with AUTHORRUNNING}\gdef\leftheadline{%
\rlap{\folio}\hfil AUTHORS suppressed due to excessive length}%
\else
\xdef\leftheadline{\rlap{\noexpand\folio}\hfil
\ignorespaces#1\unskip}%
\fi
}\let\INS=E}
\def\address#1{\bgroup\petit
\centerpar{#1}\bigskip\egroup
\catcode`\@=12
\vskip2cm\noindent\ignorespaces}
\let\INS=N%
\def@#1{\if N\INS\unskip$\,^{#1}$\else\global\footcount=#1\relax
\if E\INS\hangindent0.5\parindent\noindent\hbox
to0.5\parindent{$^{#1}$\hfil}\let\INS=Y\ignorespaces
\else\par\hangindent0.5\parindent\noindent\hbox
to0.5\parindent{$^{#1}$\hfil}\ignorespaces\fi\fi}%
\catcode`\@=12
\headline={\petit\def\newline{ }\def\fonote#1{}\ifodd\pageno
\rightheadline\else\leftheadline\fi}
\def\rightheadline{Missing CONTRIBUTION
title\hfil\llap{\folio}}
\def\leftheadline{\rlap{\folio}\hfil Missing name(s)
of the author(s)}
\nopagenumbers
\let\header=Y

\let\lasttitle=N
 \def\contribution#1{\vfill\eject
 \let\header=N\bgroup
 \textfont0=\tafontt \scriptfont0=\tafonts \scriptscriptfont0=\tafontss
 \textfont1=\tamt \scriptfont1=\tams \scriptscriptfont1=\tams
 \textfont2=\tast \scriptfont2=\tass \scriptscriptfont2=\tasss
 \par\baselineskip=16pt
     \lineskip=16pt
     \tafontt
     \raggedright
     \pretolerance=10000
     \noindent
     \centerpar{\ignorespaces#1}%
     \vskip17pt\egroup
     \nobreak
     \parindent=0pt
     \everypar={\global\parindent=1.5em
     \global\let\lasttitle=N\global\everypar={}}%
     \global\let\lasttitle=A%
     \setbox0=\hbox{\petit\def\newline{ }\def\fonote##1{}\kern2.5true
     cc\ignorespaces#1}\ifdim\wd0>\hsize
     \message{Your CONTRIBUTIONtitle exceeds the headline,
please use a short form
with CONTRIBUTIONRUNNING}\gdef\rightheadline{CONTRIBUTION title
suppressed due to excessive length\hfil\llap{\folio}}%
\else
\gdef\rightheadline{\ignorespaces#1\unskip\hfil\llap{\folio}}\fi
\catcode`\@=\active
     \ignorespaces}
\def\titlea#1#2{\if N\lasttitle\else\vskip-28pt
     \fi
     \vskip18pt plus 4pt minus4pt
     \bgroup
\textfont0=\tbfontt \scriptfont0=\tbfonts \scriptscriptfont0=\tbfontss
\textfont1=\tbmt \scriptfont1=\tbms \scriptscriptfont1=\tbmss
\textfont2=\tbst \scriptfont2=\tbss \scriptscriptfont2=\tbsss
\textfont3=\tasys \scriptfont3=\tenex \scriptscriptfont3=\tenex
     \baselineskip=16pt
     \lineskip=0pt
     \pretolerance=10000
     \noindent
     \tbfontt
     \rightskip 0pt plus 6em
     \setbox0=\vbox{\vskip23pt\def\fonote##1{}%
     \noindent
     \if>#1>\ignorespaces#2
     \else\ignorespaces#1\unskip\enspace\ignorespaces#2\fi
     \vskip18pt}%
     \dimen0=\pagetotal\advance\dimen0 by-\pageshrink
     \ifdim\dimen0<\pagegoal
     \dimen0=\ht0\advance\dimen0 by\dp0\advance\dimen0 by
     3\normalbaselineskip
     \advance\dimen0 by\pagetotal
     \ifdim\dimen0>\pagegoal\eject\fi\fi
     \noindent
     \if>#1>\ignorespaces#2
     \else\ignorespaces#1\unskip\enspace\ignorespaces#2\fi
     \vskip12pt plus4pt minus4pt\egroup
     \nobreak
     \parindent=0pt
     \everypar={\global\parindent=\oldparindent
     \global\let\lasttitle=N\global\everypar={}}%
     \global\let\lasttitle=A%
     \ignorespaces}
 \def\titleb#1#2{\if N\lasttitle\else\vskip-22pt
     \fi
     \vskip18pt plus 4pt minus4pt
     \bgroup
\textfont0=\tenbf \scriptfont0=\sevenbf \scriptscriptfont0=\fivebf
\textfont1=\tams \scriptfont1=\tamss \scriptscriptfont1=\tbmss
     \lineskip=0pt
     \pretolerance=10000
     \noindent
     \bf
     \rightskip 0pt plus 6em
     \setbox0=\vbox{\vskip23pt\def\fonote##1{}%
     \noindent
     \if>#1>\ignorespaces#2
     \else\ignorespaces#1\unskip\enspace\ignorespaces#2\fi
     \vskip10pt}%
     \dimen0=\pagetotal\advance\dimen0 by-\pageshrink
     \ifdim\dimen0<\pagegoal
     \dimen0=\ht0\advance\dimen0 by\dp0\advance\dimen0 by
     3\normalbaselineskip
     \advance\dimen0 by\pagetotal
     \ifdim\dimen0>\pagegoal\eject\fi\fi
     \noindent
     \if>#1>\ignorespaces#2
     \else\ignorespaces#1\unskip\enspace\ignorespaces#2\fi
     \vskip8pt plus4pt minus4pt\egroup
     \nobreak
     \parindent=0pt
     \everypar={\global\parindent=\oldparindent
     \global\let\lasttitle=N\global\everypar={}}%
     \global\let\lasttitle=B%
     \ignorespaces}
 \def\titlec#1{\if N\lasttitle\else\vskip-\baselineskip
     \fi
     \vskip18pt plus 4pt minus4pt
     \bgroup
\textfont0=\tenbf \scriptfont0=\sevenbf \scriptscriptfont0=\fivebf
\textfont1=\tams \scriptfont1=\tamss \scriptscriptfont1=\tbmss
     \bf
     \noindent
     \ignorespaces#1\unskip\ \egroup
     \ignorespaces}
 \def\titled#1{\if N\lasttitle\else\vskip-\baselineskip
     \fi
     \vskip12pt plus 4pt minus 4pt
     \bgroup
     \it
     \noindent
     \ignorespaces#1\unskip\ \egroup
     \ignorespaces}
\let\ts=\thinspace
\def\footnoterule{\kern-3pt\hrule width 2true cm\kern2.6pt}
\newcount\footcount \footcount=0
\def\advftncnt{\advance\footcount by1\global\footcount=\footcount}
\def\fonote#1{\advftncnt$^{\the\footcount}$\begingroup\petit
\parfillskip=0pt plus 1fil
\def\textindent##1{\hangindent0.5\oldparindent\noindent\hbox
to0.5\oldparindent{##1\hss}\ignorespaces}%
\vfootnote{$^{\the\footcount}$}{#1\vskip-9.69pt}\endgroup}
\def\item#1{\par\noindent
\hangindent6.5 mm\hangafter=0
\llap{#1\enspace}\ignorespaces}

\def\newenvironment#1#2#3#4{\long\def#1##1##2{\removelastskip
\vskip\baselineskip\noindent{#3#2\if>##1>.\else\unskip\ \ignorespaces
##1\unskip\fi\ }{#4\ignorespaces##2}\vskip\baselineskip}}
\newenvironment\lemma{Lemma}{\bf}{\it}
\newenvironment\proposition{Proposition}{\bf}{\it}
\newenvironment\theorem{Theorem}{\bf}{\it}
\newenvironment\corollary{Corollary}{\bf}{\it}
\newenvironment\example{Example}{\it}{\rm}
\newenvironment\exercise{Exercise}{\bf}{\rm}
\newenvironment\problem{Problem}{\bf}{\rm}
\newenvironment\solution{Solution}{\bf}{\rm}
\newenvironment\definition{Definition}{\bf}{\rm}
\newenvironment\note{Note}{\it}{\rm}
\newenvironment\question{Question}{\it}{\rm}
\long\def\remark#1{\removelastskip\vskip\baselineskip\noindent{\it
Remark.\ }\ignorespaces}
\long\def\proof#1{\removelastskip\vskip\baselineskip\noindent{\it
Proof\if>#1>\else\ \ignorespaces#1\fi.\ }\ignorespaces}
\def\typeset{\petit\noindent This article was processed by the author
using the \TeX\ macro package from Springer-Verlag.\par}
\outer\def\byebye{\bigskip\bigskip\typeset
\footcount=1\ifx\speciali\undefined\else
\loop\smallskip\noindent special character No\number\footcount:
\csname special\romannumeral\footcount\endcsname
\advance\footcount by 1\global\footcount=\footcount
\ifnum\footcount<11\repeat\fi
\vfill\supereject\end}

\def\12{{1\ov 2}}

\def\ov{\over}

\def\Tr{\,{\rm Tr}\,}

\contribution{Aspects of Nucleon Chiral Perturbation Theory}
\author{Ulf-G. Mei{\ss}ner}\footnote{}{\hskip -0.6truecm
Lecture delivered at the Workshop on Chiral Dynamics: Theory and Experiments,
Massachusetts Institute of Technology, Cambridge, USA, July 25 - 29,
1994, CRN 94-44}
\address{Centre de Recherches Nucl\'eaire, Physique Th\'eorique, BP 28 Cr,
F--67037 Strasbourg Cedex 2, France}
\vskip -2truecm
\titlea{1}{Introduction}
This lecture is concerned with the structure of hadrons at low
energies, where the strong coupling constant is large. Most of the
hadronic world discussed here will be made up of the light u, d (and s)
quarks since these are the constituents of the low-lying hadrons. The
best way to gain information about the strongly interacting particles
is the use of well-understood probes, such as the photon or the
massive weak gauge bosons. At very low energies, the dynamics of the
strong interactions is governed by constraints from chiral symmetry.
This leads to the use of effective field theory methods which in the
present context is called baryon chiral perturbation theory. In this
lecture, I will briefly outline the basic framework
of this effective field theory and use nucleon Compton and pion--nucleon
scattering to
discuss the strengths and limitations of it. The basic degrees of
freedom are the pseudoscalar Goldstone bosons chirally coupled to the
matter fields like e.g. the nucleons. The very low-energy face of the
low-lying baryons is therefore of hadronic nature, essentially
point-like Dirac particles surrounded by a cloud of Goldstone bosons.
Naturally, I can only cover a small fraction of the many interesting
phenomena related to low energy hadron physics. I have chosen to
mostly talk about the nucleon since after all it makes up large
chunks of the stable matter surrounding us and also is a good
intermediary between the nuclear and the high energy physicists
present at this workshop. Most of the methods presented here can easily
be applied to other problems, and as it will become obvious at many
places, we still have a long way to go to understand all the intriguing
features of the nucleon in a systematic and controled fashion. Whenever
possible, I will avoid to talk about models, with the exception of some
circumstances where they can be used to estimate some of the low--energy
constants entering the chiral perturbation theory machinery. In fact, I will
consider one of these constants and discuss to what extent we can understand
its numerical value from the so--called resonance exchange saturation picture.
Further aspects of nucleon structure related to photo- and
electropionproduction
within the framework of CHPT are discussed in V. Bernard's lecture [1].

\titlea{2}{Chiral Perturbation Theory with Nucleons}
The interactions of the strongly interacting particles at low energies are
severely constrained by the approximate chiral symmetry of the QCD Lagrangian.
This is particularly evident for the pseudoscalar Goldstone bosons which are
directly related to the spontaneous symmetry violation. In this section I will
be concerned with the inclusion the low-lying spin-1/2 baryons (the nucleons)
to the effective field theory. I will  consider the two flavor case
and mostly work in the isospin limit $m_u = m_d = \hat m$.
For a more detailed account, I refer to A. Manohar's lecture [2]. The inclusion
of such matter fields is less straightforward since these particles are not
related to the symmetry violation. However, their interactions with the
Goldstone bosons is dictated by chiral symmetry. Let us denote by $\Psi$ the
isospinor doublet including the neutron and the proton.
It is most convenient to choose a non-linear realization  of the chiral
symmetry  so that $\Psi$ transforms as $ \Psi \to K \Psi$ under $SU(2)_L \times
SU(2)_R$,
where $K$ is a complicated function that does not only depend on the group
elements $g_{L,R}$ of the $SU(2)_{L,R}$ but also on the Goldstone boson fields
collected in $U(x)$, i.e. $K(x) = K( g_L , g_R , U(x) )$ defines a local
transformation. Expanding $K$ in powers of the Goldstone boson fields, one
realizes that a chiral transformation is linked to absorption or emission of
pions (which was the theme in the days of "current algebra"
techniques). Let us restrict the
discussion to processes with one incoming and one outgoing baryon, such as
$\pi N$ scattering, pion photo- and electroproduction  or nucleon  Compton
scattering (otherwise, we would have to add contact $n$-fermion terms with $n
\ge 4$). In that case, the underlying effective Lagrangian formulated in
terms of the asymptotically observed fields takes the form
$${\cal L}_{\pi N}^{(1)} = \bar{\Psi} \biggl(i \gamma^\mu D_\mu - m
+ {1 \over 2} g_A \gamma^\mu \gamma_5 u_\mu \biggr) \Psi \eqno(2.1)$$
with $m$ the nucleon mass (in the chiral limit), $u_\mu = i u^\dagger
\nabla_\mu
U u^\dagger$, $u = \sqrt{U}$ and $D_\mu \, (\nabla_\mu)$ the chiral
covariant derivative acting on the nucleons (pions). Finally, $g_A$ is the
axial-vector coupling constant measured in neutron $\beta$-deacy, $g_A =
1.26$. Notice
that the lowest order effective Lagrangian contains one derivative and
therefore is of dimension one as indicated by the superscript '(1)'. In
contrast to the meson sector, ${\cal L}_{\pi \pi}^{(2,4,\ldots)}$,
 odd powers of the  small momentum $q$ are
allowed (thus, to leading order, no quark mass insertion appears since ${\hat
m} \sim q^2$). It is instructive to expand (2.1) in powers of the Goldstone
and external fields. From the vectorial term, one gets the minimal
photon-baryon coupling, the two-Goldstone seagull (Weinberg term) and many
others. Expansion of the axial-vectors leads to the pseudovector meson-baryon
coupling, the celebrated Kroll-Rudermann term and much more. Calculating tree
diagrams based on (2.1) leads to the current algebra results. This
is, however, not sufficient. First, tree diagrams are always real (i.e.
unitarity is violated)  and second,
the Goldstone nature of the pions can lead to large (non-analytic) corrections.
Therefore, one has to include loop diagrams making use of the chiral power
counting first spelled out by Weinberg [3] for the meson sector. In the
presence of baryons, the loop expansion is more complicated. First, since odd
powers in $q$ are allowed, a one-loop calculation of order $q^3$ involves
contact terms of dimension two and three, i.e. combinations of zero or one
quark mass insertions with zero to three derivatives. These terms are
collected in ${\cal L}_{\pi N}^{(2,3)}$ and a complete list of them can be
found in Krause's paper [4] (for the case of SU(3)). Second, the finiteness of
the nucleon mass in the chiral limit and the fact
that its value is comparable to the chiral symmetry breaking
scale $\Lambda \sim M_\rho$ complicates the low energy structure. This has
been discussed in detail by Gasser et al. [5]. Let me just give one
illustrative example. The one loop contribution  to the nucleon mass not only
gives the celebrated non-analytic contribution proportional to $M_\pi^3 \sim
{\hat m}^{3/2}$ but also an infinite shift of $m$ which has to be
compensated by a counterterm of dimension zero. It is a general feature that
loops produce analytic contributions at orders below what one would naively
expect (e.g. below $q^3$ from one loop diagrams). Therefore, in a CHPT
calculation involving baryons one has to worry more about higher order
contributions than it is the case in the meson sector. There is one way of
curing this problem, namely to go into the extreme non-relativistic limit [6]
and consider the baryons as very heavy (static) sources. Then, by a clever
definition of velocity-dependent fields, one can eliminate the baryon mass
term from the lowest order effective Lagrangian and expand all interaction
vertices and baryon propagators in increasing powers of $1 / m$.
To be specific, one writes (I follow here ref.[7])
$$ \Psi (x) = \exp[-im v\cdot x] ( H(x) + h(x) )   \eqno(2.2)$$
where $H(x)$ and $h(x)$ are velocity--eigenstates (remember that a
non--relativistic nucleon has a good four--velocity $v_\mu$) and then
eliminates
the "small" component $h(x)$. This is
similar to a Foldy-Wouthuysen transformation known from QED.
The lowest order effective Lagrangian takes the form
$${\cal L}_{\pi N}^{(1)} = \bar{H} ( i v^\mu D_\mu +  g_A S^\mu u_\mu ) H
\eqno(2.3)$$
with $S_\mu$ the covariant spin--vector (\`a la Pauli-Lubanski). In this
limit one recovers a consistent derivative expansion since the troublesome
mass term has been shifted into a string of interaction vertices. A lucid
discussion of the chiral counting rules in the presence of heavy
baryons can be found in ref.[8]. For example,
 the one loop contribution of the Goldstone bosons to the baryon
self-energy is nothing but the non-analytic $M_\phi^3$ ($\phi = \pi, K, \eta)$
terms
together with three contact terms from ${\cal L}_{\rm MB}^{(2)}$ (in SU(3)).
 However, one has to be somewhat careful still. The
essence of the heavy mass formalism is that one works with old-fashioned
time-ordered perturbation theory. So one has to watch out for the appearance
of possible small energy denominators (infrared singularities). This problem
has been addressed by Weinberg [9] in his discussion about the nature of the
nuclear forces. The dangerous diagrams are the ones  were cutting one pion
line (this only concerns pions which are not in the asymptotic in- or
out-states)
separates the diagram into two disconnected pieces (one therefore speaks
of reducible diagrams). These diagrams should be inserted in a Schr\"odinger
equation or a relativistic generalization thereof with the irreducible ones
entering as a potential. So the full CHPT machinery is applied to the
irreducible diagrams. This should be kept in mind. For the purposes I am
discussing, we do not need to worry about these complications. Being aware of
them, it is then straightforward to apply baryon CHPT to many nuclear and
particle physics problems [10-14]. I will illustrate this on two particular
examples in the next sections. Before doing that, however, I would like to
stress that most calculations are only in their infancy. It is believed
that for a good quantitative description one has to perform  systematic
calculations to order $q^4$, i.e. beyond next-to-leading order, as I will
discuss in the context of nucleon Compton scattering. A systematic analysis
to this order in the chiral expansion is not yet available. In Manohar's
lecture [2,15], an alternative approach of including the low--lying spin-3/2
decuplet in the effective field theory is discussed (based on phenomenological
considerations supplemented with some arguments from the large $N_c$ world).
In that fashion, one sums up a certain subset of graphs starting at order
$q^4$. A critical discussion of this approach can e.g. be found in ref.[16].

\titlea{3}{Nucleon Compton Scattering}
Consider low-energy (real) photons scattering off a proton, $\gamma (k) +
p(p_1) \to \gamma(k') + p(p_2)$ in the gauge $ \epsilon_0 =
 {\epsilon'}_0 =0$ (with $\epsilon_\mu$ denoting the polarization vector
of the incoming photon). In the cm-system we have $k_0 = {k'}_0 = \omega$
 and the
invariant momentum transfer squared is $t = (k - k' )^2 = -2 \omega^2 (1 - \cos
\theta )$. The $T$--matrix takes the form
$$ T = e^2 \sum_{i=1}^6 A_i (\omega , t) \, {\cal O}_i      \eqno(2.4)$$
in the operator basis of ref.[17]. Under crossing ($\omega \to -\omega$) the
$A_{1,2}$ are even and the $A_{3,4,5,6}$ are odd.  Furthermore, below the
single pion production threshold, $\omega_{\rm thr} = M_\pi$, the $A_i$ are
real. Clearly, the nucleon structure is encoded in these invariant functions.
With them at hand, one can readily calculate the differential cross
section and polarisation observables like the parallell asymmetry ${\cal
A}_\parallel$ (polarized photons scatter on polarized protons with the proton
 spin (anti)parallel to the photon direction) or the perpendicular asymmetry
${\cal A}_\perp$ (with the proton spin perpendicular to the photon direction)
(explicit formulae are given in ref.[14]).
In forward direction, the scattering amplitude takes the form
$${1 \over 4 \pi}T (\omega) = f_1(\omega^2)\,\,{\vec  \epsilon} \, '^* \cdot
{\vec \epsilon}   + i \omega \, f_2(\omega^2)\,\,
{\vec \sigma}\cdot ({\vec \epsilon} \, '^* \times {\vec \epsilon} \, ) \quad .
\eqno(2.5)$$
The energy expansion of the spin-independent amplitude $f_1 (\omega^2 )$ reads
  $$f_1(\omega^2) = - {e^2Z^2\over 4 \pi  m} + (\bar \alpha +\bar \beta)
 \,\omega^2  +  {\cal O}(\omega^4) \eqno(2.6)$$
where the first energy-independent term is nothing but the Thomson amplitude
mandated by gauge invariance. Therefore, to leading order, the photon  only
probes some global properties like the mass or electric charge of the spin-1/2
target. At next-to-leading order, the non-perturbative structure is
parametrized by two constants, the so-called electric and magnetic
polarizabilities. To lowest order, $q^3$, these are given by a few loop
diagrams, i.e. they belong to the rare class of observables free of
low--energy constants. The lowest order results [18]
$$\bar{\alpha}_p = \bar{\alpha}_n = 10 \bar{\beta}_p = 10 \bar{\beta}_n
= {5 e^2 g_A^2 \over 384 \pi^2
F_\pi^2}{1 \over M_\pi} = 13.6 \cdot 10^{-4} \, {\rm fm}^3  \eqno(2.7)$$
already describe the two main features of the data, namely that (a) the neutron
and the proton behave essentially as (induced) electric dipoles and that (b)
$(\bar \alpha + \bar \beta)_p \simeq (\bar \alpha + \bar \beta)_n$
(see e.g. the contributions by
Nathan and Bergstrom [19]). A few remarks concerning the results (2.7)
are in order. In the chiral limit of vanishing pion mass,
$\bar{\alpha}_{p,n}$ and $\bar{\beta}_{p,n}$ diverge as 1/$M_\pi$.
This is expected since the two
photons probe the long-ranged pion cloud, i.e. there is no more Yukawa
suppression as in the case for a finite pion mass. Furthermore,
a well-known  dispersion sum rule relates $(\bar \alpha + \bar \beta)$
 to the total  nucleon photoabsorption cross section.
The latter is, of course, also well-behaved in the chiral limit which at first
sight seems to be at variance with the behaviour of the expansion of the
scattering amplitude. But be aware that the general form of (2.6) has been
derived under the assumption that there is a well defined low-energy limit.
However, as has been pointed out by many
[20], the strong magnetic (M1) $N \Delta \gamma$ transition leads to a
potentially large $\Delta$ conntribution, $\bar{\beta}_{p,n}^\Delta
 \simeq 10 \cdot
10^{-4}$ fm$^3$. From the CHPT point of view, such contributions start at order
$q^4$ since they are $\sim F_{\mu \nu} F^{\mu \nu}$ (with $F_{\mu \nu}$ the
canonical photon field strength tensor which counts as $q^2$).  This problem
was addressed in a systematic fashion in refs.[21], where {\it all} terms of
${\cal O}(q^4)$ were considered (not only some as in previous works). These
new terms fall into two categories. The first one consists of
 one loop diagrams with
excatly one insertion from ${\cal L}_{\pi N}^{(2)}$. The corrresponding
low--energy constants $c_{1,2,3}$ can be estimated from resonance exchange or
determined from data on elastic $\pi N$ scattering (as discussed in section 4
and 5). The second class are genuine new counter terms from
${\cal L}_{\pi N}^{(4)}$, their coefficients could only be estimated making use
of the resonance saturation principle (which works well in the meson sector
[22]). I will come back to this in section 5.  The pertinent results for the
electromagnetic polarizabilities take the generic form [21]
$$ (\bar \alpha , \bar \beta)_{p,n} = {C_1 \over M_\pi} + C_2 \, \ln M_\pi +
C_3
\eqno(2.8)$$
where the constant $C_1$ can be read off from eq.(2.7). The loops of order
$q^4$ contribute to the second and third term whereas the large local $\Delta$
contribution enters prominently in $C_3$. Including  the theoretical
uncertainties in estimating the corresponding low--energy constants and also
the possible contributions from loops involving strangeness, one arrives at the
following theoretical predictions:
$$ \bar{\alpha}_p = 10.5 \pm 2.0 , \, \, \bar{\alpha}_n = 13.4 \pm 1.5 ,
\, \, \bar{\beta}_p = 3.5
\pm 3.6 , \, \, \bar{\beta}_n = 7.8 \pm 3.6 ,             \eqno(2.9)$$
in units of $10^{-4}$ fm$^3$. These agree (with the exception of
$\bar{\beta}_n$)
very well with the data. The two main lessons learned from this improved
calculation are: (1) The chiral expansion for electric polarizabilities
converges quickly and (2) in the case of $\bar{\beta}_p$,
the coefficient $C_2$ is
large so that the $\ln M_\pi$ term cancels most of the large and positive
$\Delta$ contribution. This is a novel effect which goes in the right direction
and shows once more that one has to include all terms at a given order.
However, since there are large cancellations  in the predictions for the
magnetic polarizabilities, one would like to see the result of a $q^5$
calculation. On the experimental side, it would be of importance to perform
independent measurements of the electric and magnetic polarizabilities to
(a) test the dispersion sum rule and (b) to lower the uncertainties in the
individual polarizabilities (these are considerably larger than the usually
quoted ones if one does not impose the constraint from the sum rule).

The spin-dependent amplitude $f_2 (\omega^2 )$ has an expansion analogous to
(2.6),
  $$f_2(\omega^2) = f_2(0) + \gamma\, \omega^2 + {\cal O}(\omega^4)
\eqno(2.10)$$
with the Taylor coefficient  $f_2 (0)$ given by
celebrated LET due to Low, Gell-Mann  and Goldberger [23],
$f_2 (0) = -(e^2 \kappa^2 )/ ( 8 \pi m^2)$,
with  $\kappa$ denoting the anomalous magnetic moment of the particle the
photon scatters off. In CHPT, $\kappa$ does not appear in the lowest order
effective Lagrangian but is given by loops and counter terms from ${\cal
L}_{\pi N}^{(2,3)}$ (this is frequently overlooked). The physics of the
so--called "spin--dependent" polarizability $\gamma$ is discussed in some
detail in refs.[7,24]. Here, I just want to point out that the LEGS
collaboration at Brookhaven intents to measure this interesting nucleon
structure constant [25]. Also, in ref.[26] the interesting observation was
made that the multipole predictions for the nucleon spin--polarizability and
for the so--called Drell--Hearn--Gerasimov sum rule are incompatible. This
again points towards the importance of independent experimental determinations
of these quantities.

\titlea{4}{Topics in Pion--Nucleon Scattering}

In this section, I will mostly discuss the chiral corrections to the S--wave
$\pi N$ scattering lengths and give some necessary definitions for the
following section. Consider first the S--wave scattering of pions off a nucleon
at rest in forward direction,
$$T^{ba} = T^+ (\omega ) \delta^{ba} + T^- (\omega ) i \epsilon^{bac} \tau^c
\eqno(2.11)$$
with $a(b)$ the isospin index of the incoming (outgoing) pion and $\omega =
v \cdot q = q_0$ denotes the pion cms energy. Under crossing, the functions
$T^\pm $ behave as $T^\pm (\omega) = \pm T^\pm (- \omega)$. At threshold,
$\omega_{\rm thr} = M_\pi$ (remember that I work to lowest order in the $1/m$
expansion), the amplitude is given by its scattering lenghts,
$$a^\pm = {1 \over 4 \pi} {1 \over 1 + \mu} \, T^\pm (\omega_{\rm thr}) \quad .
\eqno(2.12)$$
These are related to the also often used $a_{1/2}$ and $a_{3/2}$  via
$a_{1/2} = a^+ + 2 a^-$ and $a_{3/2} = a^+ - a^-$, respectively.
For the later discussion, we also need the so--called axial polarizability. For
that, consider $T^+$ not longer in forward direction and subtract the nucleon
born terms (as indicated by the overbar),
$${\bar T}^+ (\omega, \vec q, \vec q \, ' )
= t_0 (\omega ) + t_1 (\omega) \vec q \, ' \cdot \vec
q + \ldots \eqno(2.13)$$
with the kinematics $\omega = \nu = v \cdot q = v \cdot q'$ and $t = (q - q')^2
= 2 (M_\pi^2 - \omega^2 + \vec q \, ' \cdot \vec q \, )$.
The axial polarizability is  then defined via
$$\alpha_A = 2 c_{01}^+ \equiv t_1 (0)     \eqno(2.14)$$
where for completeness I have also given the relation to the
low--energy expansion parameter $c_{01}^+$ commonly used in the $\pi N$
community.

One of the most splendid successes of current algebra in the sixties was
Weinberg's prediction [27] of the S--wave $\pi N$ scattering lengths,
$$a^- = {M_\pi \over 8 \pi F_\pi^2} = 8.8 \cdot 10^{-2} / M_\pi , \quad a^+ = 0
\eqno(2.15)$$
in good agreement with the data, $a^- = 9.2 \pm 0.2$ and $a^+ = -0.8 \pm 0.4$
(in units of $10^{-2}/M_\pi$) [28]. I should stress
that in view of the confused situation about low--energy $\pi N$ scattering,
these scattering lenghts certainly should be assigned much larger
uncertainties [29]. For the sake of the argument, I will however stick to the
Karlsruhe--Helsinki values [28]. Of course, one has to worry whether the
chiral corrections will spoil this remarkable agreement.  In ref.[30], this
question was addressed. Besides the canonical one--loop diagrams, one has to
include three finite contact terms from ${\cal L}_{\pi N}^{(2)}$, which due
to crossing contribute to $T^+$,
$${\cal L}_{\pi N}^{(2)} = c_1 {\bar H}H \Tr (\chi_+) + \bigl( c_2 - {g_A^2
\over 8 m} \bigr) {\bar H} (v \cdot u)^2 H + c_3 {\bar H} u \cdot u H
\eqno(2.16)$$
but in fact, only the combination $C \equiv c_2 + c_3 - 2c_1 + g_A^2/8m$
enters the result for $a^+$.
The isospin--odd amplitude $T^-$ has to be renormalized
via a combination of four scale-dependent counter terms, $b^r (\lambda) =
b_1^r (\lambda) + b_2^r (\lambda) + b_3^r (\lambda) - 2b_4^r (\lambda)$ (here,
$\lambda$ is the scale of dimensional regularization)
with the corresponding contact terms from ${\cal L}_{\pi N}^{(3)}$ given in
[30] together with $b_4 {\bar H} [ \chi_- , v \cdot u] H$ (which was omitted in
that paper, the conclusions and numbers, however, remain unchanged). Due
to crossing, ${\cal L}_{\pi N}^{(3)}$ terms contribute to $a^-$. Defining
$L \equiv M_\pi / 8 \pi F_\pi^2$,  one arrives at
$$\eqalign{
a^- & = L \biggl[1 - \mu - \mu^2 ( 1+ {g_A^2 \over 4}) \biggr] +{L^2 M_\pi
\over
\pi} \biggl( 1 - 2 \ln {M_\pi \over \lambda} \biggr) - 64 \pi L^2 M_\pi F_\pi
b^r (\lambda) \cr
a^+ & = 32 \pi F_\pi^2 L^2 \, C \, (1 +\mu) + {3 \over 4} g_A^2 L^2 M_\pi \cr}
\eqno(2.17)$$
which shows that the exact knowledge of the low--energy constants is much
more important for $a^+$ than for $a^-$ because in the latter case their
contribution is suppressed with respect to the leading term by two powers of
$M_\pi$. To get a handle on the numerical values for $c_{1,2,3}$ and
$b_{1,2,3,4}$, the following procedure was used in ref.[30]. While $c_1$
is uniquely fixed form the pion--nucleon $\sigma$--term [7], the other
low--energy constants were estimated from resonance exchange. In this case,
one has contributions from the $\Delta$, the Roper and also from scalar
exchange. The quality of this procedure will be discussed in section 5.

Let me first consider the result for the isospin--odd scattering length. One
finds
$$a^- = (8.76 + 0.40) \cdot 10^{-2} / M_\pi = 9.16 \cdot 10^{-2} / M_\pi
\eqno(2.18)$$
which shows that the chiral corrections of order $M_\pi^2$ and $M_\pi^3$ are
small (approximately 5$\%$ of the lowest order result)
and move the prediction closer to the empirical value. Furthermore, the
dependence of this result on the actual value of $b^r (\lambda)$ is very weak.
Matters are different for $a^+$. Here, the contact terms play a prominent role
and the chiral prediction is very sensitive to the choice of certain resonance
parameters, one related to the scalar exchange and the other to the $\Delta$
contribution (for a more detailed discussion, see ref.[30]). Therefore,
 in the absence
of more stringent bounds on these parameters one can only draw the conclusion
that the chiral prediction for $a^+$ is within the empirical bounds for
reasonable values of the resonance parameters. Also, while for $a^-$ the
convergence in $\mu = M_\pi / m$ is rapid, it is much slower in the case of
$a^+$. This indicates that one should perform a $q^4$ calculation as it was
done in case for the nucleon polarizabilities  discussed in section 3.

\titlea{5}{Anatomy of a Low--Energy Constant}

To get an idea about the quality of the resonance saturation principle used in
the previous sections, I will consider here the low--energy constant $c_3$
defined in eq.(2.16). First, however, let me briefly review the underlying
idea of estimating low--energy constants from resonance exchange [22]. As the
starting point, consider meson resonances (M = V,A,S,P) chirally coupled to the
Goldstone fields collected in $U$ and the matter fields ($N$) plus baryonic
excitations ($N^*$) and integrate out the meson and nucleon resonances
$$  \exp i \int \, dx \, {\cal L}_{\rm eff}[U,N] = \int [dM] [dN^*]
 \exp i \int \, dx \,\tilde{{\cal L}}_{\rm eff}  [U,M,N,N^*] \eqno(2.19)$$
so that one is left with a
string of higher dimensional operators contributing to ${\cal L}_{\rm eff}
[U,N]$ in a manifestly chirally invariant manner and with coefficients given
entirely in terms of resonance masses and coupling constants of the resonance
fields to the Goldstone bosons. In the meson sector, i.e. considering neither
baryons nor their exciations, this scheme works remarkably well,
$$L_i = \sum_{M=V,A,S,P} L_i^M + L_i (\lambda )                 \eqno(2.20)$$
where the scale--dependent remainder $ L_i (\lambda )$ can be neglected if one
choses $\lambda$ to take a value in the resoance region (say between 500 MeV
and 1 GeV in the meson sector). But what about the baryon sector? To get an
idea, I calculate the axial polarizability defined in eq.(2.14). This amounts
to the evaluation of 5 finite one loop diagrams (and their crossed partners)
plus the contact term contribution proportional to $c_3$,
$$\alpha_A = -{2 c_3 \over F_\pi^2} - {g_A^2 M_\pi \over 8 \pi F_\pi^4} \biggl(
{77 \over 48 } + g_A^2 \biggr) = 2.28 \pm 0.10 \, M_\pi^{-3}   \eqno(2.21)$$
using the central value given in [31] but enlarging the uncertainty by a factor
2.5 (as was suggested to me by M. Sainio [32]). This amounts to
$$c_3 = -5.2 \pm 0.2 \, \, {\rm GeV}^{-1}           \quad .   \eqno(2.22)$$
In the resonance exchange picture, the dominant contributions to $c_3$ stem
from intermediate $\Delta$'s and scalar exchange with a small correction from
excitation of the Roper resonance. Varying the corresponding couplings within
their allowed values leads to
$$\eqalign{ c_3^{\rm res} & = c_3^\Delta + c_3^{N^*} + c_3^S \cr
& = (-2.5 \dots -3.2) + (-0.1 \ldots -0.2) + (-1.0 \ldots -1.6) \,
{\rm GeV}^{-1} \cr & = -3.6 \ldots -5.0 \, {\rm GeV}^{-1} \cr}  \eqno(2.23)$$
at the scale $\lambda = m_\Delta$. Comparison of (2.23) with (2.22) reveals
that
at least for this particular low--energy constant, the resonance exchange
saturation principle seems to work. Clearly, a more systematic analysis has to
be performed to draw a final conclusion. However, it is also mandatory to get
more high precision low--energy data, at present there are just too few of
these
to determine all low--energy constants up--to--and--including order $q^3$ (or
higher) and compare with predictions from resonance exchange.

\titlea{6}{An amazingly accurate QCD Prediction}

The structure of the nucleon as probed by the weak charged currents is encoded
in two form factors, the axial and pseudosclar ones,
$$<N(p')|A_\mu^a |N(p)> = \bar{u}(p') \biggl[ \gamma_\mu G_A (t) + {(p'-p)_\mu
\over 2m} G_P (t) \biggr] \gamma_5 {\tau^a \over 2} u(p)    \eqno(2.24)$$
with $t = (p' -p)^2$ and $A_\mu^a = q \gamma_\mu \gamma_5 (\tau^a /2) q$ the
isovector axial current. The axial form factor $G_A(t)$ is discussed in
V. Bernard's  lecture [1] so I will here concentrate on the induced
pseudoscalar form factor $G_P (t)$ as measured e.g. in muon capture,
 $\mu^- + p \to
\nu_\mu +n$, i.e. at $t = -0.88 M_\mu^2 \simeq -0.5 M_\pi^2$. One defines the
induced pseudoscalar coupling constant $g_P$ via
$$ g_P = {M_\mu \over 2m} G_P (t = -0.88M_\mu^2 )  \quad . \eqno(2.25)$$
The best empirical determination of $g_P= 8.7 \pm 1.6$ [33] is consistent
with the PCAC (lowest order) prediction $g_P^{\rm PCAC} = 8.9$. It is therefore
believed that a measurement of $g_P$ can test pion pole dominance but not more.
However, one can do better in baryon chiral perturbation theory. For that, one
 simply uses the chiral Ward identity
$$\partial^\mu [ \bar{q} \gamma_\mu \gamma_5 {\tau^a \over 2} q ] =
\hat m \bar q i \gamma_5 \tau^a q                            \eqno(2.26)$$
and sandwiches it between nucleon states. One arrives at [34]
$$ g_P = {2 M_\mu g_{\pi N} F_\pi \over M_\pi^2 + 0.88 M_\mu^2} - {1 \over 3}
g_A  M_\mu m r_A^2 + {\cal O}(q^2)                              \eqno(2.27)$$
with $g_{\pi N} = 13.31 \pm 0.34$ the strong pion--nucleon coupling constant
and  $r_A = 0.65 \pm 0.03$ fm the nucleon axial radius.
The relation (2.27) is known since long [35]
but its derivation solely based on the chiral Ward identity of QCD is new.
The resulting prediction is [34]
$$ g_P = (8.89 \pm 0.23) - (0.45 \pm 0.04) = 8.44 \pm 0.23       \eqno(2.28)$$
if one adds the uncertainties in quadrature. In fact, the largest uncertainty
stems from the much debated value of the pion--nucleon coupling constant.
Consequently, if one could measure $g_P$ within an accuracy of 2$\%$
(as it seems to be
feasible within present day technology [36]), one could cleanly test the QCD
versus the lowest order (PCAC) prediction. In fact, one could turn the argument
around and use such an accurate measurement to pin down the allowed range for
the strong pion--nucleon coupling constant. Finally, let me make a remark on
the form factor at other values of $t$. The recently published data on
$G_p (t)$ for $t= -0.07,-0.139$ and $-0.179$ GeV$^2$ [37] are not accurate
enough to cleanly distinguish between the lowest order and the one--loop QCD
prediction.

\titlea{7}{Concluding Remarks}

The standard model of the strong and electroweak interactions enjoys a
spectacular success, particularly at high energies. At low energies, the
symmetries help to formulate an effective field theory (EFT) which can
be used to perform precise calculations. The relevant degrees of freedom of
this EFT called chiral perturbation theory are the pseudo--Goldstone bosons
and other hadrons, but not the quarks and gluons. The pions, kaons and etas
play a special role in that they are linked directly to the spontaneous
symmetry violation QCD is believed to undergo. The exact mechanism of this
phase transition which generates the almost massless degrees of freedom is not
yet understood. In the effective Lagrangian, a whole string of terms with
increasing energy dimension is present, rendering the theory
non--renormalizable. This is, howewer, of no relevance since the EFT is
not supposed to be of use at all scales, but in the case at hand for
energies below the typical resonance masses (say 1 GeV). The power of chiral
perturbation theory stems from the observation that it is a systematic and
simultaneous expansion in small momenta (energies) and quark (pion) masses.
It is of utmost importance that to a given order one includes {\it all} terms
demanded by the symmetry requirements. This means that beyond leading order
the so--called low--energy constants enter, which are not given by the
symmetries. The finite parts of these constants have to be determined from
experiment or can be estimated from some principles like resonance exchange
saturation. While in the meson sector the machinery exists and is fully
operative, calculations in the baryon sector are hampered by the fact that
not sufficiently many accurate low--energy data exist to pin down all appearing
low--energy constants. However, with the new CW machines and the renewed
interest in low--energy domain, we will eventually leave this transitional
stage and will achieve a more satisfactory description of the effective
pion--nucleon field theory [38].
The extension to the case of three flavors is also
only in its infancy since the small parameter $M_K / 4 \pi F_\pi \simeq 0.4$
is not that small whereas in the two--flavor sector we deal with $M_\pi /
4 \pi F_\pi \simeq 0.1$. Furthermore, the closeness of the spin-3/2 decuplet
has triggered some speculations that one should include these degrees of
freedom in the EFT. Again, a systematic investigation of this approach is
not yet available, so that  at present one can not draw a final conclusion
on it. To summarize, let me emphasis that {\it low--energy hadron physics is
as interesting as any other field in physics and that exciting times are
ahead of us}. Many challenging problems, both theoretical and
experimental, remain to be solved.

\titlea{8}{Acknowledgements}

I would like to thank the organizers for
their invitation and making such an interesting workshop
possible. The work reported here
has been done in collaboration with V. Bernard, N. Kaiser and A. Schmidt, to
whom I express my gratitude.

\begref{References}{[MT1]}
\refno{[1]}V. Bernard, these proceedings
\refno{[2]}A.V. Manohar, these proceedings
\refno{[3]}S. Weinberg, {\it Physica} {\bf 96A} (1979) 327
\refno{[4]}A. Krause, {\it Helv. Phys. Acta\/} {\bf 63} (1990) 3
\refno{[5]}J. Gasser, M.E. Sainio and A. ${\check {\rm S}}$varc,
{\it Nucl. Phys.\/} {\bf B307} (1988) 779
\refno{[6]}E. Jenkins and A.V. Manohar, {\it Phys. Lett.\/} {\bf B255} (1991)
558
\refno{[7]}
V. Bernard, N. Kaiser, J. Kambor and Ulf-G. Mei{\ss}ner, {\it Nucl. Phys.\/}
{\bf B388} (1992) 315
\refno{[8]}G. Ecker, {\it Czech. J. Phys.\/} {\bf 44} (1994) 405
\refno{[9]}S. Weinberg, {\it Nucl. Phys.\/} {\bf B363} (1991) 3
\refno{[10]}E. Jenkins and A.V. Manohar, in "Effective Field Theories of the
Standard Model", ed. Ulf--G. Mei{\ss}ner, World Scientific, Singapore, 1992
\refno{[11]}Ulf-G. Mei{\ss}ner, {\it Int. J. Mod. Phys.} {\bf E1} (1992) 561
\refno{[12]}Ulf-G. Mei{\ss}ner, {\it Rep. Prog. Phys.} {\bf 56} (1993) 903
\refno{[13]}T.-S. Park, D.-P. Min and M. Rho, {\it Phys. Rep.}
{\bf 233} (1993) 341
\refno{[14]}V. Bernard, Ulf-G. Mei{\ss}ner, N. Kaiser and A. Wirzba,
"Chiral Symmetry in Nuclear Physics", in preparation
\refno{[15]}E. Jenkins and A.V. Manohar, {\it Phys. Lett.\/} {\bf B259} (1991)
353
\refno{[16]} V. Bernard, N. Kaiser and Ulf-G. Mei{\ss}ner,
{\it Z. Phys.\/} {\bf C60} (1993) 111
\refno{[17]}A.C. Hearn and E. Leader, {\it Phys. Rev.\/} {\bf 126} (1962) 789
\refno{[18]}
V. Bernard, N. Kaiser and Ulf-G. Mei{\ss}ner, {\it Phys. Rev. Lett.\/}
{\bf 67} (1991) 1515; {\it Nucl. Phys.\/} {\bf 373} (1992) 364
\refno{[19]}J. Bergstrom, A. Nathan, these proceedings
\refno{[20]}M.N. Butler and M.J. Savage, {\it Phys. Lett.\/} {\bf B294} (1992)
369; A. L'vov, {\it Phys. Lett.\/} {\bf B304} (1993) 29
\refno{[21]} V. Bernard, N. Kaiser, Ulf-G. Mei{\ss}ner and A. Schmidt,
{\it Phys. Lett.\/} {\bf B319} (1993) 269;
{\it Z. Phys.\/} {\bf A348} (1994) 317
\refno{[22]}G. Ecker, J. Gasser, A. Pich and E. de Rafael,
{\it Nucl. Phys.\/} {\bf B321} (1989) 311;
J. F. Donoghue, C. Ramirez and G. Valencia, {\it Phys. Rev.\/} {\bf D39}
(1989) 1947
\refno{[23]} F.E. Low, {\it Phys.  Rev.\/} {\bf 96} (1954) 1428;
M. Gell-Mann and M.L. Goldberger, {\it Phys. Rev.\/} {\bf 96} (1954) 1433
\refno{[24]}
Ulf-G. Mei{\ss}ner, in "Substructures of Matter as Revealed with Electroweak
Probes", eds. L. Mathelitsch and W. Plessas, Springer, Berlin, 1994
\refno{[25]}A.M. Sandorfi, private communication
\refno{[26]}A.M. Sandorfi et al., Brookhaven preprint BNL--60616, 1994
\refno{[27]}S. Weinberg, {\it Phys. Rev. Lett.\/} {\bf 17} (1966) 616
\refno{[28]}R. Koch, {\it Nucl. Phys.\/} {\bf A448} (1986) 707
\refno{[29]}G. H\"ohler, M.E. Sainio, these proceedings
\refno{[30]} V. Bernard, N. Kaiser and Ulf-G. Mei{\ss}ner,
{\it  Phys. Lett.\/} {\bf B319} (1993) 269
\refno{[31]}
G. H\"ohler, in Landolt--B\"ornstein, vol.9 b2, ed. H. Schopper,
 Springer, Berlin, 1983
\refno{[32]}M.E. Sainio, private communication
\refno{[33]}
G. Bardin et al., {\it  Phys. Lett.\/} {\bf B104} (1981) 320
\refno{[34]} V. Bernard, N. Kaiser and Ulf-G. Mei{\ss}ner, {\it Phys. Rev.}
{\bf D} (1994), in print
\refno{[35]}S.L. Adler and Y. Dothan, {\it Phys. Rev.\/} {\bf 151} (1966) 1267;
L. Wolfenstein, in: "High--Energy Physics and Nuclear Structure", ed. S.
Devons,
Plenum, New York, 1970
\refno{[36]}D. Taqqu, talk presented at the International Workshop on "Large
Experiments at Low Energy Hadron Machines", PSI, Switzerland, April 1994
\refno{[37]}S. Choi et al., {\it Phys. Rev. Lett.\/} {\bf 71} (1993) 3927
\refno{[38]}G. Ecker, {\it Phys. Lett.\/} {\bf B} (1994), in print
\endref
\byebye